\documentclass[useAMS,usenatbib]{mn2e}

\usepackage{graphics}

\title[Cluster Lensing of QSOs]{Cluster Lensing of QSOs as a Probe of $\Lambda$CDM and Dark Energy Cosmologies}
\author[Ana M. Lopes and Lance Miller]
{Ana M. Lopes \thanks{aml@astro.ox.ac.uk} and Lance Miller
\\
Department of Physics, Oxford University,
Denys Wilkinson Building, Keble Road, Oxford OX1 3RH, U.K.\\
}

\begin{document}

\date{DRAFT \today}

\pagerange{\pageref{firstpage}--\pageref{lastpage}} \pubyear{2002}

\maketitle

\label{firstpage}

\begin{abstract}
Wide-separation lensed QSOs measure the mass function and evolution of
massive galaxy clusters, in a similar way to the cluster mass function 
deduced from X-ray--selected samples 
or statistical measurements of the Sunyaev--Zeldovich effect. 
We compute probabilities of strong lensing of QSOs by galaxy clusters in 
dark energy cosmologies using semianalytical modelling and explore the 
sensitivity of the method to various input parameters and assumptions. 
We highlight the
importance of considering both the variation of halo properties with 
mass, redshift and cosmology and the effect of cosmic scatter in halo concentration. 
We then investigate 
the extent to which observational surveys for wide-separation lensed QSOs may 
be used to measure cosmological parameters 
such as the fractional matter density $\Omega_{M}$, the $\it{rms}$ linear 
density fluctuation in spheres of $8$\,h$^{-1}$\,Mpc, $\sigma_{8}$, and the dark 
energy equation of state parameter $w$.  

We find that wide-separation lensed QSOs can measure $\sigma_{8}$ and 
$\Omega_{M}$ in an equivalent manner to other methods such 
as cluster abundance studies and cosmic shear measurements. In assessing 
whether lensing statistics can distinguish between values
of $w$, we conclude that at present the uncertainty in the calibration of
$\sigma_8$ in quintessence models dominates the conclusions reached. 
Nonetheless, lensing searches based on current QSO surveys such as the Two-degree 
Field and the Sloan Digital Sky Survey with 
\mbox{$10^{4}$--$10^{5}$} QSOs should detect systems with angular separations
$\Delta\theta > 5''$ 
and hence can provide an important test of the standard cosmological model 
that is complementary to  measurements of cosmic microwave background anisotropies. 
\end{abstract}


\section{Introduction}
The first year Wilkinson Microwave Anisotropy Probe (WMAP)
measurements \citep{spergel03} of the cosmic microwave 
background
anisotropy seem to confirm that the universe is flat and that there is
a sufficient density of ``dark energy'' to accelerate the expansion of the
universe. It is not yet clear whether the dark energy is a cosmological
constant which is spatially uniform and constant with time, or instead
a dynamic cosmic field, so-called quintessence, which varies in space
and time. A combination of the WMAP observations with other astronomical 
datasets has provided a good measurement of the dark energy equation of
state parameter $w<-0.78$ ($95\%$ confidence limit)
\citep{spergel03}. However, the combination of datasets in this way means 
that independent tests of both the values of cosmological parameters and the 
underlying assumptions cannot be made. Statistical tests based on the 
abundance of massive structures can be an alternative test of the standard 
cosmological model in which structures have grown with the expansion of the 
Universe from initially Gaussian primordial fluctuations. 
Observational signatures of clusters at $z \loa 3$ contrast with observations 
of the cosmic microwave background anisotropies at $z \approx 1000$, 
forming a complementary cosmological probe. One well-known example of 
such a test is the statistics of strong lensing events.

This paper investigates the use of strong lensing statistics as probes of 
cosmology, and in particular discusses the joint
constraints that may be obtained on the equation of state of the dark
energy, $P = w\rho$ (where in this paper $w$ is assumed constant), and
on the fractional matter density $\Omega_M$ and power spectrum
normalisation, parameterised by $\sigma_8$, the $\it{rms}$ linear density 
fluctuation in spheres of $8$\,h$^{-1}$\,Mpc.
Multiple images of distant QSOs can be produced by strong lensing by 
intervening
mass overdensities such as galaxies and clusters of galaxies.
The work presented in this paper is focused on wide-separation lensed QSOs, 
$\Delta\theta \goa 5''$. These systems are produced 
by structures with the mass of clusters of galaxies, so wide-separation
lensed QSOs can probe the number and evolution of clusters in a
similar way to the cluster mass function deduced from X-ray--selected 
samples or statistical measurements of the Sunyaev--Zeldovich
effect. However, these latter methods depend on emission from, or
scattering by, the baryons in a cluster, while gravitational lensing
directly probes its mass. But, despite the existence of arcminute-separated
multiple images of background galaxies (\citealt{smail95,kneib96,sahu98}),
the detection of wide-separation lensed QSOs has
proved to be a hard task. Only two systems are known to have separations 
$\Delta\theta>5''$: $Q0957+561$, with $\Delta\theta=6.26''$ and redshift $z=1.41$,
is the widest-separation confirmed lensed QSO known \citep{walsh} and
$RX\,J0921+4529$, with $\Delta\theta=6.97''$ and $z=1.65$,
remains a wide-separation lens candidate \citep{munoz}. Other
systematic lensing searches \citep{ofek,maoz97,phillips1,phillips2}
have failed to find wide-separation lensed QSOs.
Ongoing redshift surveys such as the Two-Degree Field QSO Redshift Survey 
(2QZ)
and the Sloan Digital Sky Survey (SDSS) with 
\mbox{$10^{4}$--$10^{5}$} QSOs might be able to detect wide
separation lensed systems.  \cite{miller02} have reported evidence for
QSO lensed candidates with arcminute separations selected from 2QZ, but 
further
observational work is required to demonstrate that those
candidates are indeed strongly lensed systems. 

Strong lensing statistics have been calculated for $\Lambda$CDM
models by \cite{LiO} and for QCDM models by \cite{sarbu}. The work 
presented here not 
only extends the work performed by \cite{sarbu} to wide-separation
lenses but also uses what we consider to be more accurate
assumptions about the halo mass function and the mass distribution within
halos, and investigates more explicitly the
effect of making differing model assumptions.
First, we adopt the Sheth $\&$ Tormen \citep {sheth} halo mass
function, which should more accurately predict the mass function
of massive clusters. Second, we utilise the dark matter halo concentration
prescription of \citet{eke01} and take into account the variation of
the concentration with redshift, mass and cosmology. Here we have
explicitly considered the dependence of the concentration on $w$,
$\Omega_{M}$ and $\sigma_{8}$. For comparison with previous work,
\citet{sarbu} used the
\citet{bul01} relation and only considered the mass and redshift
dependence of concentration. Third, we
include the effects of scatter in halo concentration \citep{bul01} which was
neglected by \citet{LiO} and \citet{sarbu}. These last points are
crucial as lensing depends very strongly on halo concentration, in that
more concentrated halos are more efficient lenses.  We also note that
the statistics of strong lensing events depends critically on the assumed
value of $\sigma_8$, and by assuming an empirical relationship between 
$\sigma_8$
and $w$ the {\em cause} of variation in lensing statistics with 
cosmological parameters can become obscured.  In fact the \citet{sarbu}
results are dominated by the assumed $\sigma_8$--$w$ relationship.
In this paper we present results as a function of $\sigma_8$, $w$ and
$\Omega_M$ so that the inter-relationships between these variables may be 
understood.
We also show how calculation of the amplification bias changes when
widely separated multiple images are being considered.  

Hence, in section \ref{sec2} we describe how the
lensing probability arising from massive clusters 
may be estimated and discuss the elements that 
play a role in that
computation, in section \ref{sec3} we show the results of that analysis
and in section \ref{sec4} we discuss the outlook for being able to constrain
cosmological parameters from such measurements.

\section{The Lensing Probability}
\label{sec2}
If we define the lens system with a lens at redshift $z$, a source at redshift $z_{S}$, which is in our case a QSO, the probability for the source being lensed with an image separation greater than $\Delta \theta$ is:
\begin{equation}
P(z_{S}, >\Delta \theta) = \int_{0}^{z_{S}} (1+z)^{3} \frac{dr}{dz}\,dz \, \int_{M_{min}}^{\infty}
\frac{dn}{dM}(M,z)\,dM \, \sigma_{lens}
\label{prob}
\end{equation}
where $\sigma_{lens}$ is the lensing cross-section which is specified in
section \ref{lsec}, $n$ is the number of lenses per comoving volume,
$M_{min}$ is the minimum mass required to produce an image splitting $\Delta\theta$, $r$ is the proper
cosmological distance to the lens
and $(1+z)^{3}$ accounts for the fact that $n$ is in units of comoving
volume.  The basic ingredients that take part in the calculation of
the lensing probability are therefore the halo mass function, the
background cosmology and the lens cross-section which depends
critically on the halo profile and on the magnification bias. The
following subsections will detail these ingredients.

\subsection{The Cosmological Model}
We consider spatially flat cosmological models with a dark energy
equation of state $P_{Q}=w\rho_{Q}$, where $P_{Q}$ is the pressure, $\rho_{Q}$ is density and $w$, the so-called dark energy equation of state parameter, is assumed to be constant with time. In this case the local energy conservation
law (see e.g \citealt{pr02}) is $\dot{\rho}_{Q}=-3\left(\ddot{a}/a\right)\rho_{Q}\left(1+w\right)$, where $\rho_{Q} \propto a^{-3(1+w)}$ and $a$ is the cosmic expansion factor. Dark energy has a number of
implications in cosmology and some of these are of direct relevance
for strong lensing. Two clear consequences are the effects on the
cosmic volume and the growth of structure. For a flat model with fixed
fractional matter density $\Omega_{M}$ (and therefore $\Omega_{Q} =
1-\Omega_{M}$) the cosmic volume per unit redshift decreases as $w$
increases whereas structures start forming earlier if $w$ increases
\citep{bart02a,klypin}.  For our purposes we need to compute angular diameter
distances. In flat dark energy cosmologies the angular diameter
distance from an object at redshift $z_{1}$ to an object at redshift
$z_{2}$ is given in units of $c/H_{0}$ by:
\begin{equation}
D_{z_{1},z_{2}}=\frac{1}{(1+z_{2})}\int_{z_{1}}^{z_{2}}
\frac{dz}{(\Omega_{M}(1+z)^{3}+\Omega_{Q}(1+z)^{3(1+w)})^{0.5}}. 
\end{equation}
Other consequences of dark energy such as its influence on the mass
power spectrum, the linear and non-linear overdensity for collapse and
the dark matter halo concentrations will be discussed in later sections.  
We adopt a value for Hubble's constant, $H_{0} = 70$\,km\,s$^{-1}$\,Mpc$^{-1}$.

\subsection{The Halo Mass Function}
In this paper we consider lensing only by massive galaxy clusters, and neglect
any contribution from individual galaxies.  The results obtained should
be valid on image separation scales where the calculated lensing probability
significantly exceeds that due to individual galaxies, and from our analysis we estimate the minimum scale to be $\Delta\theta \sim 5''$. To calculate the number density and redshift distribution
of massive clusters 
we adopt the Sheth $\&$ Tormen \citep{sheth} mass function.
This is a modified version of that obtained
by the Press-Schechter formalism \citep{ps74} that fits 
better the space density of high-mass
halos determined from N-body simulations. This increase in precision
is obtained principally by introducing an additional factor into
the critical linear overdensity for collapse, where that factor is
adjusted so that the mass function matches the results from those
simulations.  This mass function produces results almost indistinguishable from
the mass function of \citet{jenkins}, which is also a fit to N-body
simulations. Both fitting formulae predict significantly greater numbers
of high-mass halos than Press-Schechter.  In adopting this function, we
implicitly have to assume that the function is valid for dark energy models
as indicated by the simulations of \citet{klypin}.  
The mass function is primarily a single-parameter modification of Press-Schechter:
the latter 
makes no assumption about background cosmology, and already fits
remarkably well to results from N-body simulations. We can be confident that
the mass function prescription will remain valid for a wide range 
of cosmological models, although we shall argue later that it is important
to test not only the mass function but also the distribution of halo
concentration values by N-body simulation (e.g. \citealt{klypin}).

We also adopt the linear growth and the linear matter power spectrum for
QCDM models obtained by \cite{ma}. The major difference between the
power spectra of the $\Lambda$CDM and QCDM models is that dark energy
may cluster spatially on scales corresponding to wavenumbers
$k \loa 0.01 h$\,Mpc$^{-1}$, whereas the cosmological constant remains
spatially smooth on all length scales. This signifies that the QCDM
power spectrum on cluster scales has the same shape as the $\Lambda$CDM model 
but has a different large-scale shape, and, especially if the normalisation
is determined from large-scale CMB measurements, 
the overall amplitude is changed \citep{ma}.  Here we use the
\cite{bardeen} fitting formula for the $\Lambda$CDM model transfer
function with the modification of \citet{sug} in order to account for the
baryons.  The resulting $\Lambda$CDM power spectra are extremely close
to those computed by \citet{eshu}.

To avoid varying too many parameters, we fix the primordial power 
spectrum index to be $n=1$ and the 
baryon density to be the WMAP best-fit value of $\Omega_{b}=0.047$.  
In fact, varying the primordial index does not change significantly the halo mass
function if it is normalised to a particular value of $\sigma_8$.
Modifying the power spectrum to include the effects of baryons in fact
{\em decreases} the lensing probability, somewhat contrary to our initial
expectations: this point will be discussed again in the context of 
halo concentrations. 

\begin{figure}
\resizebox{80mm}{!}{
\rotatebox{-90}{
\includegraphics{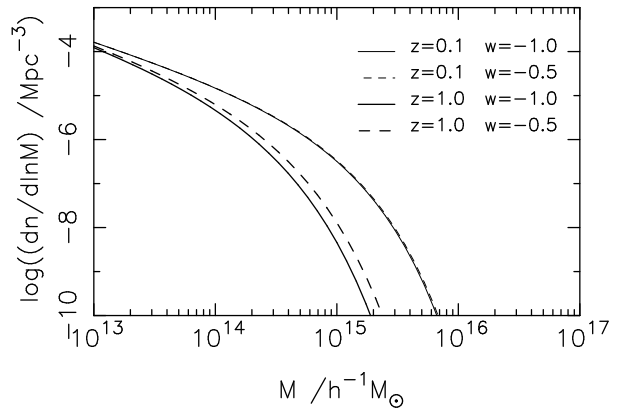}}}
\caption{The Sheth \& Tormen mass function and its (lack of)
sensitivity to the dark
energy equation of state parameter $w$. Thin lines are for $z=0.1$ and
thick lines are for $z=1.0$. Solid lines correspond to a cosmological
constant model with $w=-1.0$ and dashed lines correspond to a
quintessence model with $w=-0.5$. The fractional matter density and
the power spectrum normalisation were set to $\Omega_{M}=0.3$ and
$\sigma_{8}=0.9$ respectively.}
\label{pswz}
\end{figure}

\begin{figure}
\resizebox{80mm}{!}{
\rotatebox{-90}{
\includegraphics{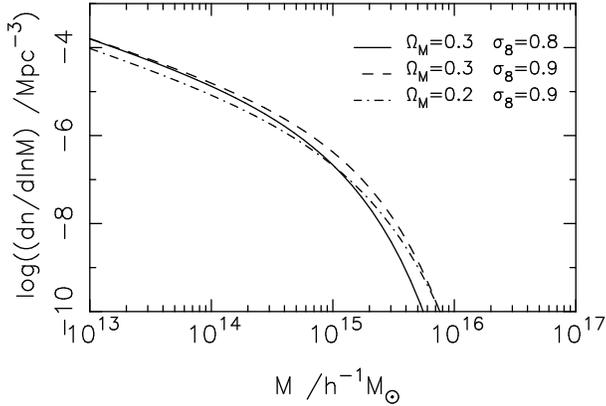}}}
\caption{The Sheth \& Tormen mass function and its sensitivity to
$\Omega_{M}$ and $\sigma_{8}$. The solid line is for
$(\sigma_{8},\Omega_{M})=(0.8,0.3)$, the dashed line for 
$(\sigma_{8},\Omega_{M})=(0.9,0.3)$ and the dash-dotted line for 
$(\sigma_{8},\Omega_{M})=(0.9,0.2)$. Here, the mass function is
computed at $z=0$ and we assume $w=-1.0$.}
\label{pss8om}
\end{figure}

One critical quantity that varies with $w$
is the linear theory critical threshold for
collapse of a halo, which is a key part of the Press-Schechter and
Sheth-Tormen prescriptions. We adopt the fitting function proposed by
\citet{wk02} which was modelled after \citet{ks96}: in fact the value of the linear 
theory critical threshold for the collapse at the typical redshift of the lens ($z \sim 0.5$)
does not vary considerably, not more than approximately $5$ per cent, over a wide range of $\Omega_{M}$ and $w$.  Figs~\ref{pswz} and \ref{pss8om} present the
Sheth $\&$ Tormen mass function as a function of halo mass. As is well
known, the mass function is a steep function of mass, decreases
with redshift and increases exponentially with $\sigma_{8}$. Fig.~\ref{pss8om} 
shows the dependence of the mass function on $\sigma_{8}$
and $\Omega_{M}$.  Fig.~\ref{pswz} plots the mass function at two
redshifts, $z= 0.1$ and $z=1$ for a cosmological model with a
cosmological constant and for a model with a dark energy equation of
state parameter $w=-0.5$. Clearly, the mass function is not very
sensitive to the dark energy equation of state parameter at $z=0.1$
but the difference between the two models is greater at higher redshifts,
and the model with $w=-0.5$ predicts more halos
than the cosmological constant model, for a fixed value of $\sigma_8$,
as shown also by \citet{klypin}.

\subsection{The Lensing Cross Section}
\label{lsec}
The lensing cross section, which may be defined as the area on the 
lens plane for which multiple imaging occurs, is given by:
\begin{equation}
\sigma_{lens}=D_{L}^{2}\int_{0}^{\beta_{crit}(z,M)}2\;\pi\,\beta\;A(z,M,\beta)\;d\beta
\label{lenscross}
\end{equation}
where $D_{L}$ is the angular diameter distance from the observer to
the lens, $\beta$ is the angle between the lens and the source,
$\beta_{crit}(z,M)$ is the critical angle for multiple imaging
and $A(z,M,\beta)$ is the amplification bias. Note here that the
amplification bias is part of the integrand and is therefore
explicitly computed as a function of the angle $\beta$ instead of
simply calculating the cross section as $\sigma_{lens}=\pi
(D_{L}\beta_{crit})^{2}$ as in previous studies.

\subsubsection{The Halo Profile}
Following previous lensing probability calculations, 
we assume that the lenses have a circularly symmetric surface mass density, $\Sigma({\bf r})=\Sigma(\left| {\bf r}\right|)$, where ${ \bf r}$ is a two-dimensional position vector in the lens plane. In this case, the lens equation reduces to a one-dimensional form \citep{maoz97,schneider92}:
\begin{equation}
\beta = \frac{D_{LS}}{D_{S}}\alpha(\theta)-\theta
\end{equation}
where $\beta$ is the angle between the lens and the source, $D_{LS}$ is the angular diameter distance between the lens and the source, $D_{S}$ is the angular diameter distance between the observer and the source, $\theta$ is the angle between the lens and the lensed image formed and $\alpha$ is the deflection angle:
\begin{equation}
\alpha(r) = \frac{4 G M (<r)}{c^{2} r}
\end{equation}
where $M (<r)$ is the projected mass enclosed within radius $r$ and is given by:
\begin{equation}
M (<r ) = 2 \pi \int_{0}^{r} \Sigma(r')r'dr'.
\end{equation}
We discuss the possible effects of departures from smooth, circularly-symmetric 
lenses in section 4.

The density profile of the lenses is modelled with the NFW profile \citep{nfw} which 
appears to be a good fit to numerically simulated halos over a wide range of 
masses in various cosmological scenarios. The density profile is: 
\begin{equation}
\rho(r) = \frac{\rho_{S}}{\frac{r}{r_{S}}(1+\frac{r}{r_{S}})^{2}} 
\end{equation}
where $\rho_{S}$ and $r_{S}$ are the characteristic density scale and
scale radius, respectively. This profile is steeper than the singular
isothermal sphere at large radii but is flatter at smaller
radii. \citet{moore} simulations prefer a steeper inner profile 
although the observational evidence for the formation of radial
arcs in clusters \citep{sand} is favouring a shallower inner
profile. The inner slope is thus an issue under debate and we will not
discuss it here but we point out that the lensing probability is rather
sensitive to its value \citep{LiO}: lensing probability increases as the
inner slope steepens because for standard NFW profiles the projected mass
density is usually lower than the critical density for lensing for all
but the innermost radii of a cluster.

The surface mass density of the NFW profile is \citep{bart96}:
\begin{equation}
\Sigma(x) = \frac{2 \rho_{S} r_{S}}{x^{2}-1} f(x)
\end{equation}
where $x$ is the radial coordinate in units of scale radius \mbox{$x\equiv\frac{r}{r_{S}}$}.
As a result the bend angle takes the form:
\begin{equation}
\alpha(x) = \frac{16 \pi \rho_{S} r^{2}_{s}}{c^{2}} \frac{g(x)} {x}
\end{equation}
where g(x) is \citep{bart96}:
\begin{equation}
g(x) = \int_{0}^{x} \frac{y f(y)}{y^{2}-1}dy.
\end{equation}

In order to calculate the lensing cross section it is convenient to calculate
a critical angle for lensing, $\beta_{crit}$.
Multiple imaging occurs if and only if the angle $\beta$
satisfies $|\beta|\leq\beta_{crit}$ where $\beta_{crit}$ is the
solution of $d\beta/d\theta=0$. In the case of perfect alignment
between the observer, circularly-symmetric lens and source, $\beta=0$, and
$|\beta|<\beta_{crit}$ there are three images formed, for
$|\beta|=\beta_{crit}$ two images and for $|\beta|>
\beta_{crit}$ only one image is formed.  Non-circularly symmetric lenses
can produce more than three images.
A key question for the lensing cross section is how to calculate
the inner parameters $\rho_{s}$ and $r_{s}$ and their variation with mass, redshift and cosmological model. These parameters can be related to the virial parameters of the halo through the halo concentration parameter. The latter is basically a measurement of the relative size of the core with respect to the virial radius. Here we adopt the approach of calculating the halo concentration from prescriptions based on N-body simulations.
 
\subsubsection{Dark Matter Halo Concentrations}
Formally, the mass of the NFW profile diverges, so to define the mass we
adopt the standard formalism of calculating 
the mass within the virial radius $R_{vir}$ of a sphere with 
density $\Delta_{vir}$ times the critical density $\rho_{crit}$:
\begin{eqnarray}
M_{vir} &=& 4 \pi \int_{0}^{R_{vir}} \rho r^{2}dr = 4 \pi \rho_{S} r^{3}_{s} f(c_{vir}) \nonumber
        \\&=& \frac{4 \pi}{3} \Delta_{vir} \rho_{crit}(z) R_{vir}^{3}
\end{eqnarray}
where $f(c_{vir})=\ln(1+c_{vir})-\frac{c_{vir}}{1+c_{vir}}$ and we define the halo concentration parameter as $c_{vir}=\frac{R_{vir}}{r_{S}}$. $\Delta_{vir}$ is the non-linear overdensity at virialisation.
With these definitions we can compute the scale radius $r_{s}=R_{vir}/c_{vir}$ and 
the characteristic density $\rho_{S}$:
\begin{equation}
\rho_{s} = \frac{\Delta_{vir}}{3} \rho_{crit}(z) \frac{c_{vir}^{3}}{f(c_{vir})}.
\end{equation}
In the work that follows 
we take into account the variation of the mean halo concentration 
with redshift, halo mass and cosmology, and in
particular with $w$. The assumptions on these dependencies turn out to
make a significant difference to the overall lensing probabilities:
\cite{LiO} had no
dependence of concentration on mass in their $\Lambda$CDM analysis. 
\cite{sarbu} 
included the redshift and mass variation in their QCDM studies but
did not include any dependence on $w$, assuming instead the values
appropriate for a $\Lambda$CDM cosmology.  To evaluate those dependencies
we adopt the concentration prescription of \cite{eke01}.  That
prescription is based around the premise that halo concentration 
increases as the redshift of formation of a dark halo increases, a
reflection of the higher matter density at earlier epochs.  A 
characteristic redshift of formation that varies with the
shape of the power spectrum is calculated and a value for a single parameter 
can be 
found which yields concentration values that match well the results from
N-body simulations for a wide range of cosmological models, and has been
tested against the original NFW results for masses up to $10^{15} M_{\sun}$.
Concentrations deduced in this way seem to be generally in broad agreement
with values measured from fitting to the X-ray profiles of clusters
over the range $10^{14} \loa M_{vir} \loa 10^{16} M_{\sun}$ (e.g. \citealt{WuXue}).
It does appear that the dependence on mass in those samples may be steeper
than predicted, but X-ray selection of clusters is likely to have some
effect on the observed relationship, and 
at high masses, which are most important for wide-separation lensing, 
comparable concentration values are obtained from
the measurements and from the \citet{eke01} prescription.

In this paper we extend the
\cite{eke01} prescription in a natural way to accommodate
dark energy models working under the assumption that the
algorithms derived from numerical simulations in $\Lambda$CDM models
are also valid in QCDM models. We have used the fitting function of
\cite{wk02}, modelled after \cite{ks96} for the non-linear
overdensity at virialisation, $\Delta_{vir}$, which increases with $w$. 
This arises because structures start forming earlier
in models with a higher $w$, the mean energy in
a collapsing object is larger and hence a higher
overdensity for virialisation is required \citep{wk02}.
\begin{figure}
\resizebox{80mm}{!}{
\rotatebox{-90}{
\includegraphics{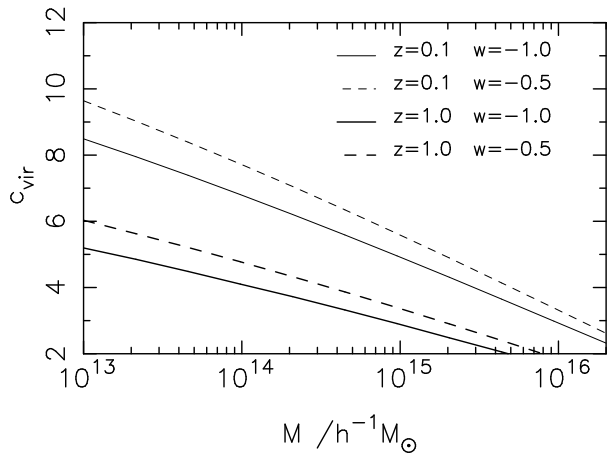}}}
\caption{Dark matter halo concentrations function for four different
combinations of redshift and dark energy equation of state parameter
$(z,w)$. The fractional matter density and the power spectrum
normalisation were set to $\Omega_{M}=0.3$ and $\sigma_{8}=0.9$
respectively. Thin lines are for $z=0.1$, thick lines for
$z=1.0$. Solid lines correspond to a $\Lambda$CDM model ($w=-1.0$) and
dashed lines correspond to a quintessence model with $w=-0.5$.}
\label{conc1}
\end{figure}
\begin{figure}
\resizebox{80mm}{!}{
\rotatebox{-90}{
\includegraphics{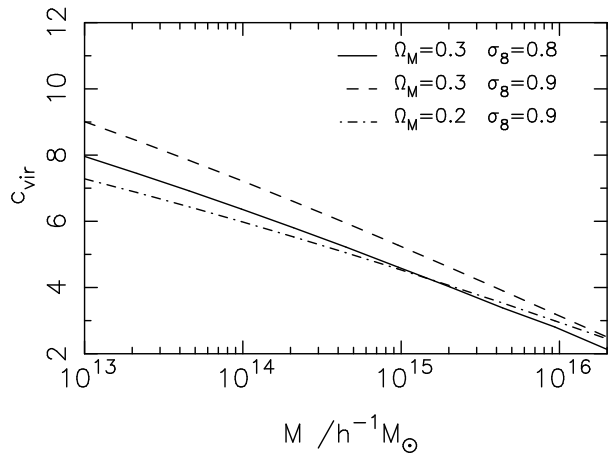}}}
\caption{Dark matter halo concentrations for three
($\sigma_{8}$,$\Omega_{M}$) combinations. The solid line is for
$(\sigma_{8},\Omega_{M})=(0.8,0.3)$, the dashed line for
$(\sigma_{8},\Omega_{M})=(0.9,0.3)$ and the dash-dotted line for
$(\sigma_{8},\Omega_{M})=(0.9,0.2)$. Concentrations were computed
at $z=0.$ and assumed a $\Lambda$CDM model.}
\label{conc2}
\end{figure}
Figures \ref{conc1} and \ref{conc2} show the concentrations obtained
for differing cosmological models. The concentrations decrease with
the mass of the halo since high halo masses are formed later in
structure formation models.  In quintessence models 
structures form earlier than in $\Lambda$CDM, resulting in higher 
concentrations for a given mass.
This trend is evident in the plot of Fig.~\ref{conc1}.
Although the \cite{eke01} prescription has not yet been exhaustively
tested for $w$ models, comparison with the dark energy simulations of
\citet{klypin} show good general agreement, with higher $w$ models
exhibiting higher concentrations for a given mass.
Fig.~\ref{conc2} presents
concentrations for three combinations of ($\sigma_{8}$,$\Omega_{M}$). An
increase in $\sigma_{8}$ makes the halo concentrations higher and a
decrease in $\Omega_{M}$ lowers the concentrations.  

One key additional factor to consider is the scatter in halo concentration
values about the mean relations computed above. The numerical
simulations show that there is a scatter in the concentration values
which may be associated with a spread in collapse epochs and/or
formation histories (\citealt{bul01,wechsler02}).  This scatter has a dramatic effect on
the wide-separation lensing probability because clusters with NFW profiles and moderate
concentration values have surface mass densities which only rise above the
critical density for lensing at small radii.  Even a 30\,percent increase
in concentration can bring the mass density above the critical value
over a substantially larger range of radii, thereby greatly increasing the
lensing probability at large separations.  So giving a minority of the
cluster halo population larger concentrations than the mean produces a
significant increase in lensing probability, as discussed in section\,3.
The scatter also reduces the differences between the various models however,
as also discussed later in this paper.  The amount of scatter measured
from N-body simulations depends
critically on which halos are chosen for study:  relaxed halos are
better fit by NFW profiles and have a moderate scatter.  \citet{bul01}
argue that the scatter appears to have
log-normal distribution with $\sigma_{c}=0.18$ and we incorporate
this distribution into the computation of the lensing probability.
Finally we note that the \citet{eke01} prescription has been tested in a range 
of power spectra shapes. Here we find that introducing $\Omega_{b}=0.047$ lowers the mean halo concentrations and therefore results in a decrease of the lensing probability.

\subsubsection{The Amplification Bias}
The effect of the amplification bias on the lensing cross section, $A(z,M,\beta)$,
is to boost the lensing probability because
sources that are intrinsically dimmer than the flux limit of the
survey $f_{lim}$ are brought into the sample by the magnification.  If
the source QSOs have a single power-law flux distribution with 
redshift-independent slope 
$\alpha$, $N_{z_{S}}(> f)=f^{-\alpha}$ then the amplification bias is
given by:
\begin{eqnarray}
A(z,M,\beta) &=& \frac{N_{z_{S}}(> f_{lim}/\mu(z,M,\beta))}{N_{z_{S}}(> f_{lim})} \nonumber \\
	     &=& \mu^{\alpha}(z,M,\beta)
\end{eqnarray}
where $\mu$ is the magnification factor which for circularly symmetric lenses is given 
by:
\begin{eqnarray}
\mu&=&\left|\frac{\theta}{\beta}\right|\left|\frac{d\theta}{d\beta}\right|
\nonumber \\
&=&\left|\frac{\theta}{\beta}\right|\left|\left(\frac{D_{LS}}{D_{S}}\frac{d\alpha(\theta)}{d\theta}-1\right)^{-1}\right|.
\end{eqnarray}
In order to calculate the magnification factor we need to solve the
lens equation and obtain the three (or two) solutions. In surveys for
wide-separation lensed systems, where the image separation is larger
than the survey resolution, the multiple images are separately
detected and we need to compute the probability that at least two
images will appear brighter than the survey flux selection limit.
Hence the relevant quantity to compute is the amplification of
the second-brightest of the multiple images. 
This approach differs from that commonly taken in which  multiple images are assumed to be unresolved 
and the amplification bias is calculated from the sum of the fluxes of the
lensed images. In addition, in this paper
we compute both the magnification factor and the minimum mass required to
produce a given image splitting as a function
of the angle $\beta$, in contrast to the usually-followed
procedure of calculating the minimum mass for lensing assuming perfect
alignment between the observer, the lens and the source.
\section{Results}
\label{sec3}
\begin{figure}
\resizebox{80mm}{!}{
\rotatebox{-90}{
\includegraphics{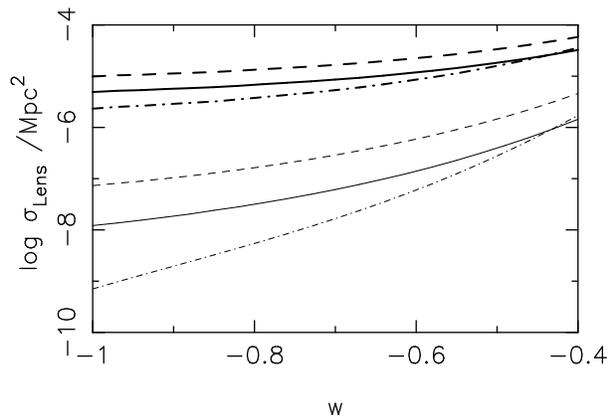}}}
\caption{The lensing cross section as a function of $w$ for a halo of mass
$M_{Halo}=10^{14}M_{\odot}$ and redshift $z=0.3$. The QSO is at
redshift $z_{QSO}=1.5$. Curves are for three different
$(\sigma_{8},\Omega_{M})$ combinations. Solid curves are for $\Omega_{M}=0.3$
and $\sigma_{8}=0.8$, dashed for $\Omega_{M}=0.3$ and
$\sigma_{8}=0.9$ and dash-dotted curves for $\Omega_{M}=0.2$ and
$\sigma_{8}=0.9$. No amplification bias has been applied. Thin lines show
the lensing cross section $\sigma_{lens}=\pi
(D_{L}\beta_{crit})^{2}$. Thick lines include the additional effect of
introducing the scatter in the halo concentrations.}
\label{cross}
\end{figure}
\begin{figure}
\resizebox{80mm}{!}{
\rotatebox{-90}{
\includegraphics{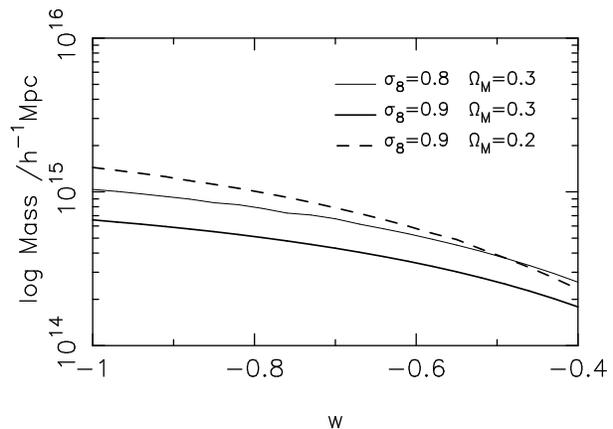}}}
\caption{The minimum mass required to lens a QSO at redshift $z_{QSO}=1.5$
with image separation $\Delta\theta = 10''$. The lens is at
redshift $z_{Lens}=0.3$ and it is assumed that there is close to perfect
alignment, $\beta \approx 0$. The minimum
mass is plotted as a function of $w$ for three pairs of
$(\sigma_{8},\Omega_{M})$. The thin solid line is for
$(\sigma_{8},\Omega_{M})=(0.8,0.3)$, the thick solid is for
$(\sigma_{8},\Omega_{M})=(0.9,0.3)$ and the dashed line is for
$(\sigma_{8},\Omega_{M})=(0.9,0.2)$.}
\label{minmass}
\end{figure}

Having explored the sensitivity of the mass function and the dark
matter halo concentrations to the dark energy equation of state
parameter as well as to the fractional matter density and the power
spectrum normalisation, we now investigate the influence of the same
cosmological parameters on the remaining lensing ingredients, such as
the lensing cross section and the minimum mass for lensing.

We first look at the lensing cross section: in order to separate the
various effects we do not at this stage
apply the magnification bias, and hence effectively calculate
$\sigma_{lens}=\pi (D_{L}\beta_{crit})^{2}$. Fig.~\ref{cross} shows
that the lensing cross section increases monotonically in quintessence
universes for fixed $\sigma_{8}$, and the amplitude of the
cross section is larger if the power spectrum normalisation is
raised. This arises because the concentrations are larger in those
cases which in turn make lensing more efficient. If the matter content
is decreased from $\Omega_{M}=0.3$ to $\Omega_{M}=0.2$ the cross
section is lowered, although the difference between the models
decreases for higher values of $w$. This occurs because halo concentrations 
are in fact lower for lower values of $\Omega_{M}$ (see Fig.~\ref{conc2}). 
Also shown is the additional effect of including a
scatter in the value of the concentrations, in particular by assuming
the log-normal distribution with $\sigma_{c}=0.18$. Including the
scatter significantly increases the lensing cross section:
for example, in a model with $(\sigma_{8},\Omega_{M})=(0.9,0.3)$ and
$w = -1$ its value is raised by a factor
of approximately one hundred, as the tail of high concentration 
halos dominates the statistics.  We should expect that for large scatter
in halo concentration the differences between cosmological models should
be reduced (assuming the scatter itself is independent of cosmology)
and this is reflected also in Fig.~\ref{cross}.  Concentration scatter was
not included in previous work by \citet{sarbu} and \citet{LiO}.

The other key
ingredient on the lensing calculation is the minimum mass required to
produce multiple imaging and this is included in formula \ref{prob} as
the lower limit of the mass integral. This is generally calculated
assuming perfect alignment between the lens, observer and the source,
that is to say with $\beta=0$. For the lensing probabilities here we
actually compute the minimum mass as a function of the angle $\beta$
but in order to analyse how its value varies with cosmological
parameters we assume $\beta=0$. Fig.~\ref{minmass} shows that the
minimum mass required to lens a QSO at redshift $z_{QSO}=1.5$ with an
image separation $\Delta\theta = 10''$ and a lens redshift
$z_{lens}=0.3$ clearly decreases with $w$ and $\sigma_{8}$. For
$\sigma_{8}=0.9$ the minimum mass required is greater for a universe
with a lower fractional matter density. 
\begin{figure}
\resizebox{80mm}{!}{
\rotatebox{-90}{
\includegraphics{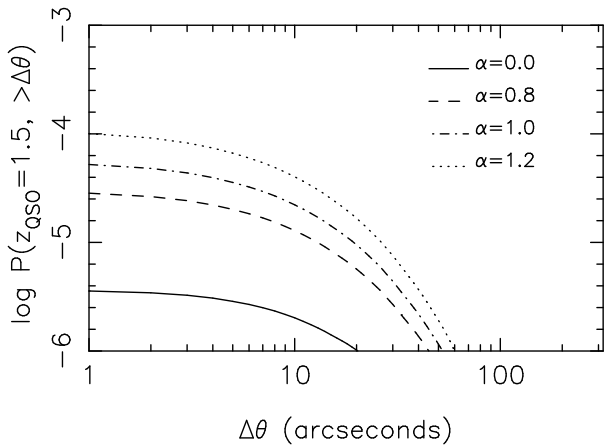}}}
\caption{The lensing probability as a function of image separation
$\Delta\theta$ for four different values of the slope of the QSO
number-counts. The solid line shows $\alpha=0$, corresponding to no
amplification bias. The fractional matter density and the
power spectrum normalisation were set to $\Omega_{M}=0.3$ and
$\sigma_{8}=0.9$ respectively. The dark energy equation of state parameter was set to $w=-1$.}
\label{slope}
\end{figure}
\begin{figure}
\resizebox{80mm}{!}{
\rotatebox{-90}{
\includegraphics{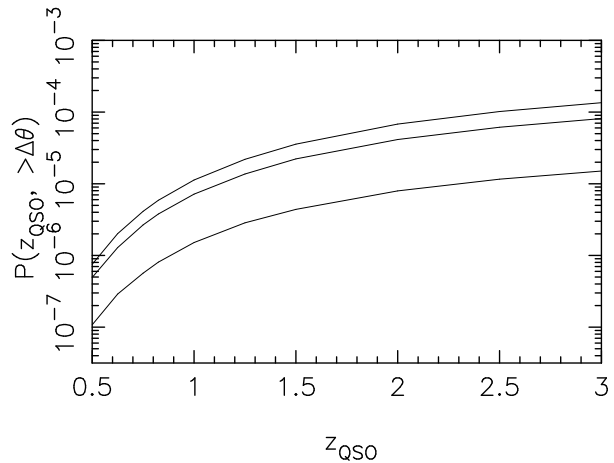}}}
\caption{The lensing probability as a function of source redshift, $z_{QSO}$, for three different image separations: $\Delta\theta=5''$ (top line), $\Delta\theta=10''$ (middle line) and $\Delta\theta=30''$ (bottom line), assuming $\Omega_{M}=0.3$, $\sigma_{8}=0.9$ 
and $w=-1$.}
\label{redeffect}
\end{figure}

We can now look at the probability distribution of image separations
(Figs~\ref{slope} and \ref{probwmap}). There is a clear and steep
fall of the probability toward higher image separations which is
a consequence of a mass function that falls off quite rapidly combined
with a higher required minimum mass for lensing events with
larger separations.  Here we underline the sensitivity of the lensing
probability to the slope of the source number-counts. Fig.~\ref{slope} 
shows the expected behaviour: a steeper slope favours
higher lensing rates. Note that if magnification bias is not accounted
for then the lensing probability is reduced by a factor of about ten. 
At faint radio and optical selection levels, with $B \goa 20$, the slope
of the QSO number-counts are significantly flatter than Euclidean,
although because we need to know the slope at flux levels fainter than
the unlensed limit for the survey being searched for lensed QSOs the
actual value of $\alpha$ is often uncertain.  The observed apparent slope
for the 2QZ optical QSO survey has value $\alpha \approx 0.8$ at faint
magnitudes, although this value does not take account of sample incompleteness.
Hereafter we shall assume a slope $\alpha=1$: to first order the change in 
lensing probability resulting from assuming a different value for $\alpha$
may be estimated from Fig.~\ref{slope}.

Assumptions about the typical source redshift and the distribution of 
source redshifts also affect the lensing
probability. Fig.~\ref{redeffect} shows that the lensing optical depth
is a steep function of source redshift: if a source is
moved from redshift $z=1.5$ to $z=2.5$, the lensing probability is
increased by approximately a factor of three. This behaviour is also
found in estimates of the lensing optical depth for giant arcs \citep{wamb}. 
Fig.~8 shows that applying an 
accurate estimate of the median source redshift is essential.  But we
also expect that a distribution of source redshifts about the median 
would increase the lensing probability, and indeed if we adopt
the redshift distribution of the 2QZ 10K catalogue \citep{croom} 
we find that the lensing probability is increased by approximately 10$\%$.

\begin{figure}
\resizebox{80mm}{!}{
\rotatebox{-90}{
\includegraphics{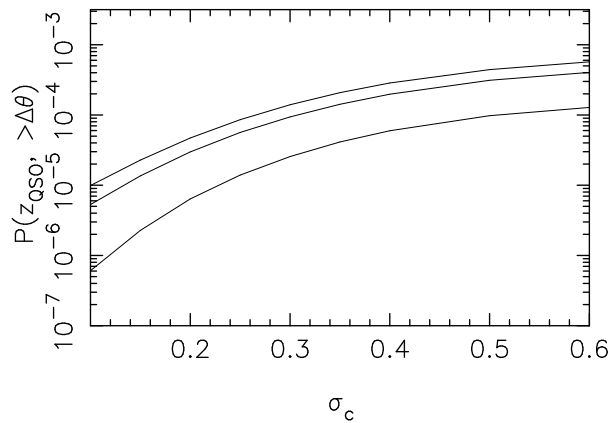}}}
\caption{The lensing probability as a function of concentration scatter, $\sigma_{c}$, for three different image separations: $\Delta\theta=5''$ (top line), $\Delta\theta=10''$ (middle line) and $\Delta\theta=30''$ (bottom line), assuming $\Omega_{M}=0.3$, $\sigma_{8}=0.9$ and $w=-1$.}
\label{scatter}
\end{figure}
Uncertainties in halo concentrations which can be expressed by $\sigma_{c}$ can also alter the lensing probability. This is shown in  Fig.~\ref{scatter}: if the scatter in halo concentrations is changed from $0.18$ (the value assumed in our analysis) to $0.4$ the lensing probability is increased by a factor of six. Therefore, an accurate value of the scatter in halo concentrations is fundamental.   
 
The effect of increasing $w > -1$ on the probability distribution of image separations
is to boost the integrated lensing probability. The difference
between the models increases for larger $w$ (Fig.~\ref{probwmap}). 
This is partly owing to halo concentrations being larger
in quintessence models, making the lensing cross section higher, and
partly owing to the minimum mass required for multiple
imaging decreasing with $w$ (Fig.~\ref{minmass}): the mass function of galaxy
clusters falls sufficiently steeply that the small decrease in mass has
a significant effect.
\begin{figure}
\resizebox{80mm}{!}{
\rotatebox{-90}{
\includegraphics{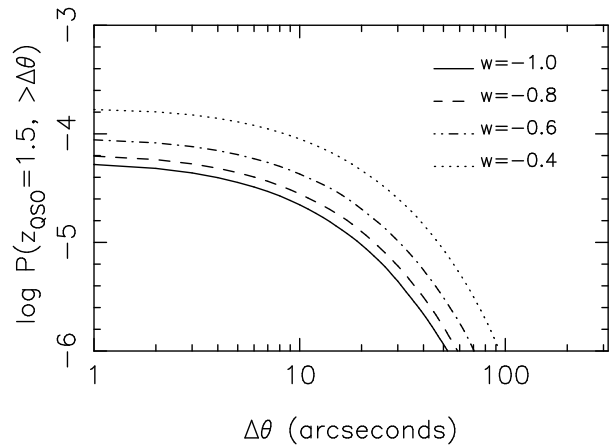}}}
\caption{The lensing probability as a function of image separation
$\Delta\theta$ for four different values of $w$ in a flat universe
with $\sigma_{8}=0.9$ and $\Omega_{M}=0.3$.}
\label{probwmap}
\end{figure}

\begin{figure}
\resizebox{80mm}{!}{
\rotatebox{-90}{
\includegraphics{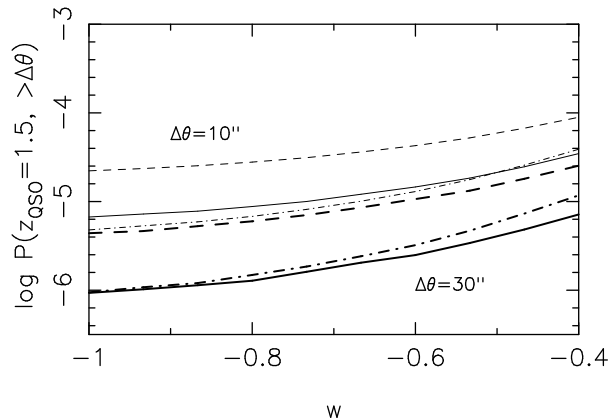}}}
\caption{The probability of a QSO at redshift $z_{QSO}=1.5$ being lensed
with an image separation bigger than $\Delta\theta$ in a quintessence
model. The sensitivity of the probability to $\sigma_{8}$ and
$\Omega_{M}$ is also shown. Solid lines are for $\Omega_{M}=0.3$ and
$\sigma_{8}=0.8$. Dashed lines are for $\Omega_{M}=0.3$ and
$\sigma_{8}=0.9$ and dash-dotted lines are for $\Omega_{M}=0.2$ and
$\sigma_{8}=0.9$. Thick lines correspond to a minimum image separation
of $\Delta\theta = 30''$ and thin lines correspond to $\Delta\theta =
10''$.}
\label{probwsep}
\end{figure}

Fig.~\ref{probwsep} explores the sensitivity of the lensing
probability to $\Omega_{M}$ and $\sigma_{8}$ for differing minimum
values of separation, $\Delta\theta=10''$ and
$\Delta\theta=30''$. The latter probability is about ten times smaller than the
former. We confirm the expected increase of probability with $w$ as in
Fig.~\ref{probwmap} and note that the probability also increases
with $\sigma_{8}$. Indeed the lensing cross section increases with
$\sigma_{8}$ and the minimum mass for lensing also decreases with this
cosmological parameter. In addition, there is an exponential rise of
the mass function with its value. The overall effect is therefore an
increase of the lensing probability. When we lower $\Omega_{M}$ from
0.3 to 0.2 the probability decreases.

\begin{figure}
\resizebox{80mm}{!}{
\rotatebox{-90}{
\includegraphics{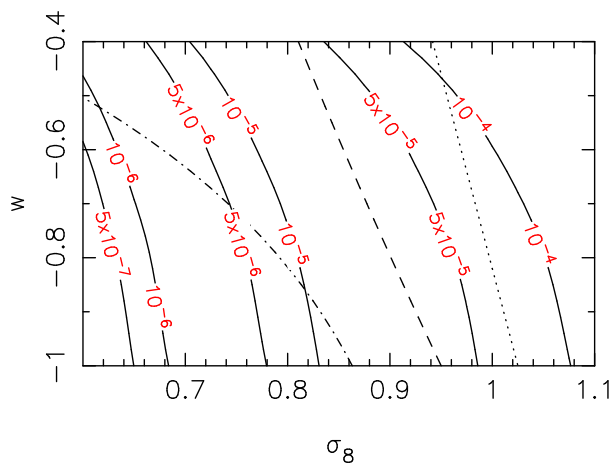}}}
\caption{Contours of probability $P(z_{QSO}=1.5,\Delta{\theta}>10'')$
in the $w-\sigma_{8}$ plane for a universe with fractional matter 
density $\Omega_{M}=0.3$. The dashed and dotted lines show the 
constraints from cluster abundances as obtained by \citet{ws} 
and \citet{lokas}, respectively. The dashed-dotted line shows the result of 
COBE power spectrum normalisation \citep{ma}.}
\label{conts8w}
\end{figure}

\begin{figure}
\resizebox{80mm}{!}{
\rotatebox{-90}{
\includegraphics{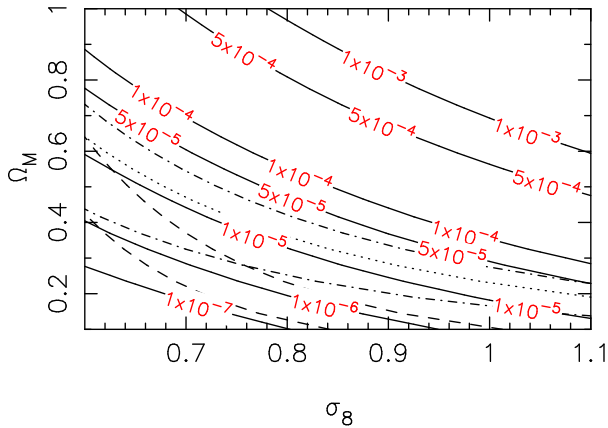}}}
\caption{Contours of probability $P(z_{QSO}=1.5,\Delta{\theta}>10'')$
in the $\sigma_{8}-\Omega_{M}$ plane for a universe dominated by a
cosmological constant ($w=-1$). The area enclosed by the dashed lines
represents the constraints from the study of the local cluster
abundance \citep{allen}. The area enclosed by the dot-dashed lines
shows the constraints obtained from the cosmic shear measurement of
\citet{hoe}. The dotted line is the WMAP relation 
\citep{spergel03}.}
\label{conts8om}
\end{figure}

There are evident cosmological parameter degeneracies and those are
clearly shown on the contour plots provided in Figs~\ref{conts8w}
and \ref{conts8om}. One has to bring in other independent cosmological
constraints in order to break the degeneracies. In Fig.~\ref{conts8w} 
we overplot the best-fit result from the 
abundance of rich galaxy clusters obtained by \citet{ws} and \citet{lokas} as well as
the constraint obtained from the COBE measurement of the cosmic
microwave background (\citealt{ma,bunn}). 
In Fig.~\ref{conts8om} we show the constraints from a study of the present-day
number density of galaxy clusters \citep{allen} and from a cosmic
shear measurement \citep{hoe}. \citet{jarvis} and \citet{pierpaoli} review the variation found in these measurements from studies performed by other groups. The WMAP best-fit result is also shown \citep{spergel03}.

\begin{figure}
\resizebox{80mm}{!}{
\rotatebox{-90}{
\includegraphics{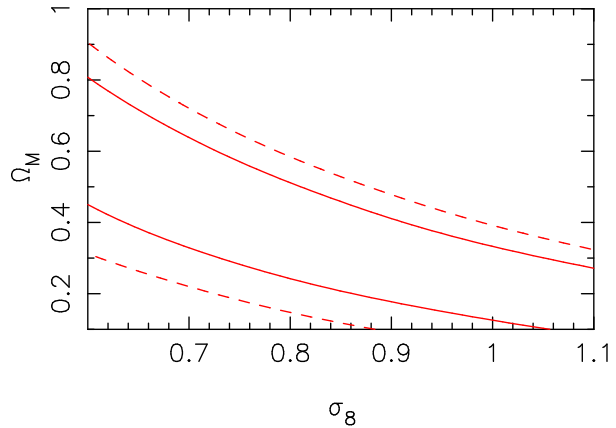}}}
\caption{The $68.3$ (solid line) and $95.4$ (dashed line) per cent confidence level contours on $\sigma_{8}$ and $\Omega_{M}$ (assuming $w=-1.$) obtained for a survey of $10^{5}$ sources, four of which are lensed with $\Delta\theta > 5''$.}
\label{conflevels}
\end{figure}

\begin{figure}
\resizebox{80mm}{!}{
\rotatebox{-90}{
\includegraphics{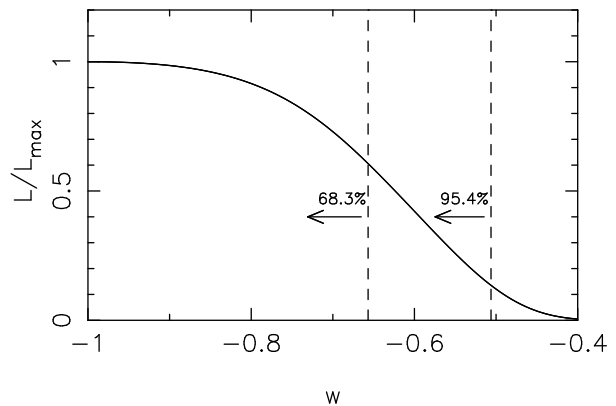}}}
\caption{The $68.3$ ($w < -0.66$) and $95.4$ ($w < -0.51$) per cent
confidence level likelihood constraints on $w$ (assuming
$\Omega_{M}=0.3$ and $\sigma_{8}=0.9$) obtained for a survey of
$10^{5}$ sources, four of which are lensed with $\Delta\theta > 5''$.}
\label{confw}
\end{figure}

The ability of surveys to actually discriminate between cosmological
models will depend on survey size and on the numbers of detected
lenses of course.  As an illustration, we assume a survey of $10^{5}$
sources containing four lensed sources with 
$\Delta\theta > 5''$, which corresponds to the model prediction for a flat
cosmological model with $w=-1$, $\sigma_{8}=0.9$ and $\Omega_{M}=0.3$,
and the source redshift distribution of the 2QZ 10K catalogue
\citep{croom}.
The likelihood function chosen depends on the number
of lenses as 
$$
L= \frac{\lambda^{N} e^{-\lambda}}{N!}
$$
where $\lambda$ is the expected number
of lensed systems and $N$ is the actual
number observed; we do not here consider information
contained in the distribution of angular separations. 
Fig.~\ref{conflevels} shows the $68.3$ and $95.4$ per cent confidence
level contours in the $\sigma_{8}-\Omega_{M}$ plane.
Fig.~\ref{confw}
shows the constraints on $w$ when one assumes $\Omega_{M}=0.3$ and
$\sigma_{8}=0.9$ (marginalising over the current uncertainty in $\sigma_8$
would significantly broaden the likelihood function). 
The confidence intervals are somewhat broader than can be
attained from current generations of cosmic microwave background,
weak lensing measurements or deductions from cluster abundance
measurements, but it is important to note that these methods are all
complementary tests of the cosmological model.  The CMB measurements
probe the Universe at $z \sim 1000$; cluster abundance determinations
are dependent on correct modelling of thermal emission from baryons in
cluster potential wells; cosmic shear measures the mean fluctuations
in matter density on Mpc scales; the abundance of strong lensing
events measures the extreme high-mass tail of the mass function, with
a small number of lensed systems being generated by the most extreme
objects.  In the final analysis the abundance of strong lensing events
may tell us more about the formation of the most massive
structures in the universe than about the values of cosmological parameters.
\section{Discussion and Conclusions}
\label{sec4}
We have calculated cluster lensing probabilities of QSOs in dark
energy models. The lensing probabilities have been explored over a wide
range of angular separations and we have taken into account the
cosmology, mass and redshift dependence of halo concentration, while
assuming that the algorithms derived from dark matter halo simulations
in $\Lambda$CDM models remain valid in QCDM models. Ultimately, these
algorithms should be tested against numerical simulations in dark
energy cosmologies, but simulations to date indicate that these
prescriptions are at least approximately correct \citep{klypin}. 
What issues remain that could still have an important effect on the predictions 
of this type of modelling?  At this stage we may be reasonably confident
that the Sheth $\&$ Tormen mass function is sufficiently well tested
that it can be considered a robust prescription, but we should be aware
that modifications to the fluctuations giving rise to cosmological
structure would lead to differing mass functions.  As one example, 
non-Gaussian initial
perturbations significantly increase the mass function at high redshifts
(\citealt{matarrese00, silk03}) and we should expect halos of a given 
mass to have formed at earlier epochs, and hence have higher values
of halo concentration.

There is currently some uncertainty about the universality of the NFW
profile and about its applicability to lensing studies.  Merely
by introducing the known scatter in halo concentration we see a significantly non-linear
effect on lensing probability. If, in turn, this scatter is greater then the value adopted in this paper ($\sigma_{c}= 0.18$) then that will have a great effect in the lensing probability. As we have seen in Fig.~\ref{scatter}, if the scatter in halo concentrations is changed from $0.18$ (the value assumed in our analysis) to $0.4$, the lensing probability is increased by a factor of six.
Variations
in the inner slope of the halo profile would also have a significant effect \citep{LiO,oguri}.
There have been claims based on N-body simulations that the inner
slope may be steeper than the standard NFW value \citep{moore}, or
that its value steepens with mass \citep{ricotti}, and there
is evidence that some clusters at least may have slopes that are flatter
than NFW \citep{sand}.  Even if there is not a systematic departure from
the canonical NFW profile, scatter about that relation could lead to
increase in lensing probability in a similar way to that introduced by
the scatter in halo concentration.  Other effects to be taken into account
include the effect of substructure, especially in recently-formed,
non-relaxed clusters: this has already been shown to be important
for the statistics of extreme giant arcs \citep{meneg}. Finally, the effect of individual
galaxies in a cluster may be noticeable on the smallest angular scales discussed in
this paper, although present indications are that this is a relatively
small effect in giant arc statistics \citep{meneg00} and, likewise, the inclusion 
of a massive cD galaxy in the centre of the cluster does not seem to 
significantly enhance the lensing cross section for giant arcs \citep{meneg2}.

We have also adopted a circularly symmetric lens profile, whereas even
relaxed individual clusters are likely to be triaxial.  Again, this 
effect has been shown to be potentially significant for the statistics
of extreme giant arcs \citep{bart02a, ogurib}, although it is not yet clear
whether there is any significant effect on the statistics of less
extreme lensing events.

In applying these models to actual surveys, we should of course also
take care to ensure that we are using accurate estimates for the 
slope of the QSO number-counts and that we take account of the redshift
distribution of QSOs. In particular, an accurate estimation of the median 
redshift of the sources is essential as the lensing probability is a steep function of source redshift. For optical QSO surveys this is known exactly and therefore does not constitute a source of uncertainty. However, for radio surveys the median redshift of the sources can be quite uncertain and therefore provide a systematic error in the lensing probability: if the median redshift is increased from $z=1$ to $z=1.5$ the lensing probability is increased by a factor about three (Fig.~\ref{redeffect}).  
  
With these caveats, we can use the models produced here to understand what
constraints, if any, may be placed on cosmological models as a result
of either detecting or not detecting wide-separation lensed QSOs in
surveys such as SDSS and 2QZ.
Within the cosmic concordance model, a flat model with
$\Omega_{M}=0.3$ and a WMAP value of $\sigma_{8}=0.9$ we find that the
lensing probability increases with the dark energy equation of state
parameter and attribute this effect to the fact that the
concentrations of dark matter halos are higher in quintessence models
and therefore lenses become more efficient. This effect
combined with a mass function that is very steep and a lower minimum
mass required for multiple imaging makes more halos available for
strong lensing. We also note that the $\sigma_{8}$--$\Omega_{M}$
degeneracy is equivalent to the degeneracy found in other methods such
as cluster abundances, cosmic shear measurements and the WMAP analysis
(Figs.~\ref{conts8om} and ~\ref{conflevels}). 

In considering whether lensing statistics can distinguish between values
of $w$, we see that at present the uncertainty in the calibration of
$\sigma_8$ dominates the conclusions reached.  Cluster-normalised
models predict that lensing probability is not a sensitive indicator
of the value of $w$, whereas COBE-normalised values of $\sigma_8$
indicate that as $w$ increases, $w > -1$, the expected number of
lensing events decreases.  Further work is needed to tie down
the $\sigma_8$ normalisation in $w$ models.  Of course, we should
not expect cluster- and CMB-determined normalisations to agree
except at a single point in parameter space that corresponds to the 
values for the universe we actually inhabit, but at present we cannot
be confident that such a point can be identified. Nonetheless, the determination of the absolute number of lenses at redshifts $z \loa 3$ should provide a good test of the $\Lambda$CDM model, independently of, and in direct contrast with, 
cosmic microwave background anisotropy measurements at  $z \approx 1000$.
Alternative methods, such as 3D weak lensing \citep{heavens} and weak
lensing tomography \citep{jain}, although at present with current
surveys are not able to constrain $w$, potentially offer a more robust
probe with an estimated accuracy for $w$ better than $5\%$ for surveys
such as the one proposed with the LSST \citep{jain}. 

For a flat cosmological constant model with $\Omega_{M}=0.3$, a WMAP
value of $\sigma_{8}=0.9$ and the source redshift distribution of the 2QZ 
10K catalogue \citep{croom}, the
modelling predicts probabilities of $4.0\times10^{-5}$,
$2.5\times10^{-5}$ and $4.7\times10^{-6}$ for lensing events with
separations larger than $5''$, $10''$ and $30''$ respectively. 
This indicates that a survey with $10^{5}$ sources will find approximately
4 lensed QSOs with separations greater than $5''$, 2 lensed QSOs with
separations greater than $10''$ and none with separations greater than
$30''$. 
For surveys such as the 2QZ and SDSS the
survey selection effects will somewhat alter these numbers. 
If {\em no} wide
separation lensed QSOs are detected that would rule out models with
high $\sigma_8$.  Conversely, if large numbers are found, or if any
lensed systems significantly larger than $1'$ are confirmed, we would
need a significant modification to the model presented here,
including either significant effects from the presence of substructure
or perhaps requiring assumptions in the standard model, such as Gaussianity,
to be relaxed.
At this stage it would be premature to speculate further.  Challenges
in the immediate future, then, are to test the expected levels of
lensing probabilities against numerical simulations, both to take
account of substructure and cluster shape, and to test the predictions in the cases of
dark energy models.  Joint comparison between multiple image
probabilities and the statistics of extreme giant arcs would be a
good simultaneous test.
\vspace*{5mm}

\noindent
{\large \bf ACKNOWLEDGEMENTS}\\
AML acknowledges the support of the Portuguese Funda\c{c}\~{a}o para a Ci\^{e}ncia e a Tecnologia.

\label{lastpage}

\end{document}